\documentclass[fleqn,10pt]{wlscirep}
\usepackage[utf8]{inputenc}
\usepackage[T1]{fontenc}
\usepackage{setspace}
\usepackage{graphicx}
\usepackage{amsmath}
\usepackage{gensymb}
\usepackage{comment}
\usepackage{psfrag}
\usepackage{float}
\usepackage[version=4]{mhchem}
\usepackage{siunitx}
\usepackage{longtable,tabularx}
\usepackage{subcaption}
\usepackage{xcolor}
\usepackage{dashrule}
\usepackage{threeparttable}
\usepackage{booktabs}
\usepackage{tikz}
\usepackage{pdflscape}
\usepackage{array}
\usepackage{hyperref} 
\usepackage{caption}
\usepackage{subcaption} 
\usepackage{tabularx}   
\usepackage{adjustbox}   
\usepackage{microtype}

\definecolor{magenta2}{RGB}{255,0,255}
\definecolor{blue2}{RGB}{0, 127, 255}
\definecolor{grey2}{RGB}{85, 85, 85}
\definecolor{teal}{RGB}{23, 190,207}
\definecolor{orange}{RGB}{255,127, 14}
\definecolor{pythonblue}{RGB}{31 ,119, 180}
\definecolor{pythongreen}{RGB}{95, 186, 125}
\definecolor{pythonyellow}{RGB}{255, 165, 0}
\tikzstyle{longdashed}=                  [dash pattern=on 6pt off 2pt]
\tikzstyle{dashdotdot}=              [dash pattern=on 4pt off 2pt on \the\pgflinewidth off 1pt on \the\pgflinewidth off 2pt]

\title{Data-driven Sensor Placement for Predictive Applications: A Correlation-Assisted Attribution Framework (CAAF)}

\author[1,*]{Sze Chai Leung}
\author[2,3]{Di Zhou}
\author[2]{H. Jane Bae}
\affil[1]{Department of Mechanical and Civil Engineering, California Institute of Technology, Pasadena, CA 91125, USA.}
\affil[2]{Lynn Booth \& Kent Kresa Department of Aerospace, California Institute of Technology, Pasadena, CA 91125, USA.}
\affil[3]{Department of Mechanical and Aerospace Engineering, University of Tennessee, Knoxville, TN 37996, USA.}

\affil[*]{sleung@caltech.edu}

\begin{abstract}
Optimal sensor placement (OSP) is critical for efficient, accurate monitoring, control, and inference in complex physical systems. We propose a machine-learning-based feature attribution (FA) framework to identify OSP for target predictions. FA quantifies input contributions to a model’s output; however, it struggles with highly correlated input data often encountered in practical applications for OSP. To address this, we propose a Correlation-Assisted Attribution Framework (CAAF), which introduces a clustering step on the candidate sensor locations before performing FA to reduce redundancy and enhance generalizability. We first illustrate the core principles of the proposed framework through a series of validation cases, then demonstrate its effectiveness in realistic dynamical systems such as structural health monitoring, airfoil lift prediction, and wall-normal velocity estimation for turbulent channel flow. The results show that the CAAF outperforms alternative approaches that typically struggle due to the presence of nonlinear dynamics, chaotic behavior, and multi-scale interactions, and enables the effective application of FA for identifying OSP in real-world environments.
\end{abstract}
\begin{document}

\flushbottom
\maketitle

\thispagestyle{empty}

\section*{Introduction}
Sensing involves the conversion of real-world physical phenomena into digital data representations. As sensing technologies and computational capabilities advance, there is a growing dependence on sensor measurements for diverse applications. However, since sensors typically provide localized observations, practical constraints on sensing resources impose fundamental limitations on data acquisition, preventing exhaustive spatial coverage or continuous temporal monitoring. Therefore, sensors must be strategically positioned to maximize information gain under these constraints. In particular, determining optimal sensor configurations for complex nonlinear systems remains a problem of significant scientific interest and complexity due to the multi-scale and multi-process nature of these systems and the strong spatiotemporal correlations inherent in their dynamics. Representative examples include turbulent flow sensing \cite{mons2017optimal,verma2020} and structural damage monitoring \cite{tibaduiza2020damage}, where the goal is to attain maximum information about the system states while minimizing the number of deployed sensors. Similarly, environmental monitoring systems require carefully designed sensor networks to ensure accurate measurements of parameters like temperature and air quality \cite{andersson2023environmental, castello2010optimal}. In autonomous systems like ground and aerial vehicles, optimal sensing enables precise environmental perception, facilitating robust decision-making and control \cite{yeong2021sensor}. The integration of machine learning (ML) further amplifies the importance of sensing, as model performance depends heavily on sensor data quality. Consequently, it is crucial to couple the ML inference process directly with sensor identification, a synergy that traditional, non-ML-based methods are unable to achieve. In this work, an optimal sensor configuration is defined as the subset of candidate sensors that maximizes the predictive accuracy for a target quantity of interest ($q$) based on sensor observations ($d$). By minimizing redundancy and maximizing information content, these configurations can improve the performance of ML models in applications such as autonomous driving and industrial automation. Thus, systematic approaches to sensor placement remain essential for balancing resource constraints against information requirements across numerous domains.

The problem of finding optimal sensor placement (OSP) has been extensively studied, leading to the development of various methods that utilize analytical or algebraic formulations to determine sensor locations \cite{kammer1991sensor,liu2018optimal,manohar2018}. The proper orthogonal decomposition with QR-pivoting (Pivoted-QR) method has been developed for OSP and demonstrated on applications such as sea surface temperature reconstruction \cite{manohar2018} and greenhouse flow\cite{tabib2023data}. Optimal placements can also be found by maximizing the system's observability for state estimation \cite{singh2005}. The effective independence (EI) method \cite{kammer1991sensor} is proposed to identify the OSP for structural health monitoring (SHM) by maximizing the determinant or trace of the Fisher information matrix and applied to the design of space stations. Viewing the sensor placement problem from the perspective of optimization, some employed Bayesian inference to optimize sensor locations for reducing parameter estimation uncertainty \cite{huan2024optimal}. Meanwhile, information-theoretic approaches exist where mutual information is utilized to identify the best sensing configurations \cite{krause2008near,otto2022}. With the recent emphasis on ML and artificial intelligence, it is worth highlighting various ML-based sensor placement schemes that leverage data-driven techniques \cite{andersson2023environmental,wang2019,zhong2023,shi2024novel}. For example, an unsupervised OSP approach relying on an attention mechanism to prune insignificant channels for SHM was recently explored by Shi et al. \cite{shi2024novel}. Furthermore, Huan and Marzouk \cite{huan2016sequential} formulated the sequential optimal experimental design problem as a dynamic program, leveraging approximate dynamic programming, a foundational reinforcement learning technique, to derive optimal sensing policies that maximize long-term information gain for a nonlinear contaminant source inversion problem.

Among these approaches, an ML-based method with significant potential for OSP is feature attribution (FA), which assigns importance scores to the input features of a model (e.g., words in a sentence, pixels in an image) based on their influence on the output, revealing the model's decision-making logic. Initially conceptualized for game theory by Shapley \cite{shapley1953}, a fair FA method should quantify the contribution of individual participants in a game based on four fundamental axioms: symmetry, efficiency, dummy, and additivity. The Shapley value, derived from these principles, was among the first to offer a systematic approach to estimating FA \cite{chen2023algorithms}. The GradientSHAP \cite{ancona2017GradientSHAP} and KernelSHAP \cite{lundberg2017,aas2021KernalSHAP} algorithms were proposed with an aim to estimate the Shapley value using different approaches. 

More recently, FA has expanded to become a prevalent technique in ML research for quantifying the contributions of input features to a model’s output \cite{sundararajan2017,sikdar2021,linardatos2020explainable,wang2020attribution}. Various algorithms have been developed with a focus on ML applications, including Saliency Map \cite{simonyan2013salientmap}, DeepLift \cite{shrikumar2017Deeplift}, and Integrated Gradients (IG) \cite{sundararajan2017}. Regardless of the chosen algorithm, FA relies on an accurate surrogate model to capture the relationship between feature inputs and the inference target. By applying an FA algorithm to a model, scores are assigned to the input features based on their significance to the target predictions. It can be used not only to eliminate redundant input features, enhancing training efficiency and accuracy, but also to explain decisions made by large models. It has notable applications in visualizing attention regions in computer vision tasks \cite{wang2020attribution} and revealing critical words or phrases in natural language processing models \cite{sikdar2021}. While FA enables researchers to analyze and compare the behaviors of ML models with human reasoning patterns, its ability to quantify input importance also provides a natural framework for sensor selection. However, its application to sensor placement remains underexplored. Unlike many existing OSP methods that rely on application-specific objective functions, data-driven FA provides a domain-agnostic framework, provided a predictive model for the quantity of interest is available. In this approach, sensor optimality is implicitly defined by the model’s objective function, such as minimizing the loss function. For a well-trained model, where this objective has been satisfied, the most significant features identified by FA represent the optimal sensor subset for maximizing predictive accuracy. Consequently, the optimal sensor configuration is identified by selecting those sensors whose measurements yield the highest attribution scores.
It requires no coupling with or modification of the model architecture, allowing direct application to existing prediction models. Therefore, extending the application of FA to sensor identification offers a promising research direction.

However, similar to many existing data-driven OSP methods, FA remains vulnerable to the challenges posed by highly correlated data commonly encountered in practical scenarios due to the violation of FA's underlying axioms. This correlation refers to the inherent inter-dependencies within physical signals, such as spatial or temporal similarity arising from governing dynamics, distinct from artificial correlation induced by measurement noise or sensor error. Correlated features cause a breakdown of the additivity axiom in Shapley attribution theory, which states that the total contribution of two features must equal the sum of their individual contributions, since part of their contribution may be shared. They complicate FA analysis as learned models may assign high importance scores to correlated inputs. This can occur because models, such as neural networks, can distribute weights arbitrarily among correlated features without losing predictive performance and disregard these inter-dependencies among input features. FA algorithms applied to such models may assign high importance scores to multiple correlated sensors, resulting in redundant sensor selections that convey overlapping information. This redundancy poses a significant limitation in OSP tasks, where the goal is to maximize information efficiency. For instance, when dealing with spatially correlated inputs, the selected sensors may cluster closely together, limiting the diversity of information captured. While such behavior is often acceptable or even desirable in the context of model interpretability, where highlighting groups of related features aids qualitative understanding, identifying optimal sensor locations necessitates scrutiny of individual elements rather than general regions. The ability to pinpoint the most informative sites is crucial.

To address this issue, we propose the Correlation-Assisted Attribution Framework (CAAF) for identifying OSP. CAAF introduces a clustering step prior to model training and FA. This clustering process groups correlated candidate sensor locations together and selects a representative point from each cluster, thereby capturing the essential information within the group. As a result, highly correlated candidates are effectively filtered out before model construction to avoid redundancy. Following this step, only these cluster centers are retained as the final set of candidates for FA. Candidate clustering is performed using a correlation metric chosen based on the characteristics of the application. This preprocessing step enhances the robustness of the framework, ensuring that the selected sensors provide diverse and non-redundant information. In summary, the CAAF identifies optimal sensor locations through five major steps. 
\begin{enumerate}
    \item Cluster the initial set of candidate sensors using a predefined correlation metric
    \item Identify the cluster centers that serve as optimal representatives for their respective clusters
    \item Develop a data-driven model to predict the target variable using inputs from the identified cluster-center sensors
    \item Apply an FA algorithm to the model to evaluate the contributions of each cluster center's inputs, and rank the cluster centers based on their significance to the model's predictions
    \item Select the desired number of cluster centers as the optimal sensor configuration based on the rankings
\end{enumerate}

In this work, we first present empirical evidence spanning from image classification tasks to synthetic validation problems to demonstrate the working principles of CAAF. These results highlight the advantages of the integrated clustering step and confirm the framework's effectiveness in idealized settings. Subsequently, we demonstrate CAAF’s applicability across three practical, control-oriented scenarios with inherent spatial or temporal correlation: identifying the most informative elements on a cantilever beam for SHM, predicting airfoil lift from surface pressure measurements, and inferring off-wall wall-normal velocity from wall pressure in wall-bounded turbulence. These problems are chosen because efficiently determining the OSP is highly desirable for predicting the quantities of interest critical to downstream control applications, yet inherently nontrivial due to the complexity of these systems. These studies illustrate the framework’s effectiveness relative to alternative methods and showcase its utility in realistic, high-dimensional datasets. Overall, the CAAF provides an ML-based approach to sensor selection for highly correlated physical systems. Its fully data-driven nature and adaptability offer a powerful paradigm for tackling complex sensor placement problems across engineering applications.

\section*{Results}


\subsection*{Correlation-aware clustering: image clustering}

\begin{figure}[t]
    \centering
    \includegraphics[width=\textwidth]{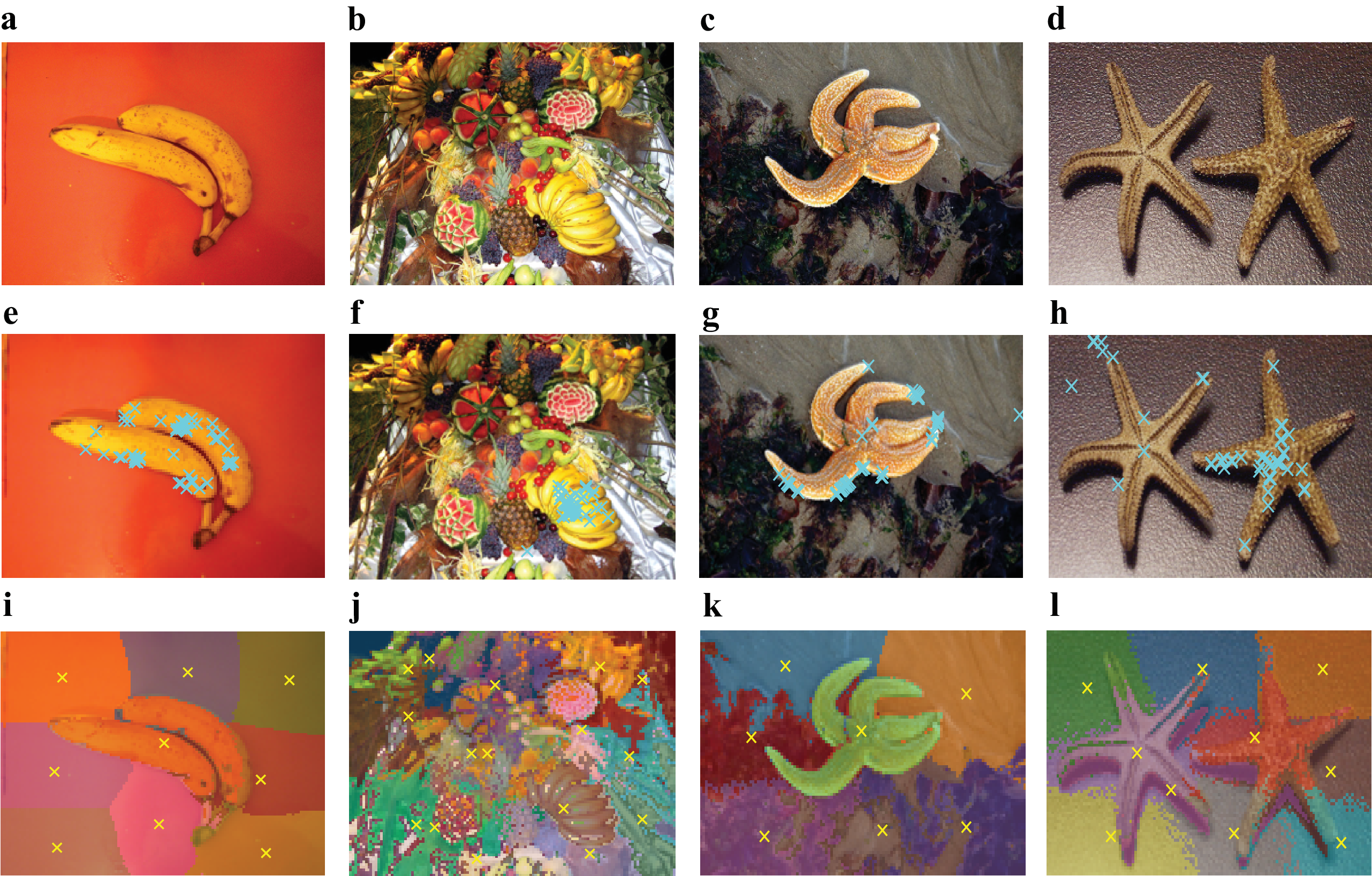}
    \caption{\textbf{Comparison of naive feature attribution (FA) and clustering-based candidate selection}. 
    \textbf{a--d} Original ImageNet samples of bananas and starfish \cite{deng2009imagenet}. 
    \textbf{e--h} Naive FA results, where cyan crosses denote the 50 pixels with the highest FA scores. 
    \textbf{i--l} Clustering-based candidate selection. Yellow crosses mark the cluster centers which define the reduced candidate set for subsequent CAAF analysis. Shaded regions indicate distinct pixel clusters.}
    \label{fig:image_cluster}
\end{figure}

FA methods have been extensively applied to explain deep ML models, with one prominent application being the interpretation of image classification models by visualizing the attention regions that overlap with objects of interest \cite{wang2020attribution, sikdar2021}. However, this tendency to highlight adjacent locations may not be ideal for sensor placement, as it often results in redundant sensors. Figure~\ref{fig:image_cluster} presents four selected samples from the ImageNet image classification dataset \cite{deng2009imagenet,imagenet_data}. If FA is directly applied to an image classification model to determine optimal color-sensor locations, the resulting sensors tend to cluster within the object of interest, as these are the pixels the model primarily uses for classification (Fig.~\ref{fig:image_cluster}e--h). Such a distribution is rarely optimal, as it lacks information diversity and may be particularly problematic for tasks that rely on global contextual information from the surrounding environment.


After implementing the clustering step prior to FA methods in image classification, the process systematically identifies all potential regions of interest along with their representative pixels as demonstrated in Fig.~\ref{fig:image_cluster}i--l. The results show distinct clusters corresponding to different objects (e.g., banana and starfish), with their respective cluster centers effectively representing multiple key regions of interest. The clustering correlation metric is defined as the Euclidean distance between pixels, integrating information from both the RGB color channels and the spatial positions of the pixels. We emphasize that this visualization represents only the intermediate clustering output, not the final CAAF results. This qualitative analysis illustrates how clustering can be used to efficiently pre-select salient image features and pinpoint the representative pixels within them. Importantly, these cluster centers form a reduced candidate set for the subsequent FA analysis in CAAF, improving computational efficiency and sensor selection performance.

\subsection*{Attribution based on correlation: correlated synthetic data}
 During the training of deep learning models, backpropagation optimizes the combined effect of features rather than individual contributions. If two features are perfectly correlated, the network may assign similar weights to both, as any linear combination of the two features yields the same output. This creates ambiguity in FA --- while the model relies on the collective signal, FA methods may misleadingly assign importance to both features equally, even though only one is theoretically necessary. Although not visually shown here, we empirically demonstrate this phenomenon by training a neural network on two identical features, which indeed results in equal FA values for both features, revealing a key limitation in directly applying FA to interpret feature importance for correlated inputs.

We now further demonstrate how CAAF effectively addresses this challenge through clustering via an idealized sensor selection problem using synthetic data, with experimental details and results presented in Table~\ref{tab:21_demo}.
The task involves optimally selecting two sensors from three candidates, where Sensors 1 and 2 exhibit high mutual correlation, while Sensor 3 is statistically independent. Row ``Corr($\boldsymbol{d}_i$, $\boldsymbol{q}$)'' in the table lists the correlation coefficients between each sensor input and the target variable, providing a reference for their expected attribution-based contributions.
If we naively apply FA without accounting for correlations, the method selects Sensors 1 and 2 due to their high individual FA scores in row ``Naive FA''. However, these sensors are highly correlated (with a correlation value of 0.9), rendering one redundant. In contrast, the CAAF first clusters the correlated candidates together, identifying Sensor 2 as the cluster center for the correlated group (Sensors 1 and 2), as shown in row ``Cluster Label''. As shown by the ``Clustered FA'' scores, when performing FA on only the cluster centers (Sensors 2 and 3), the framework correctly selects Sensors 2 and 3 as the optimal pair, aligning with intuitive expectations and avoiding redundancy.

\begin{figure}[t]
\centering
\includegraphics[width=0.4\textwidth]{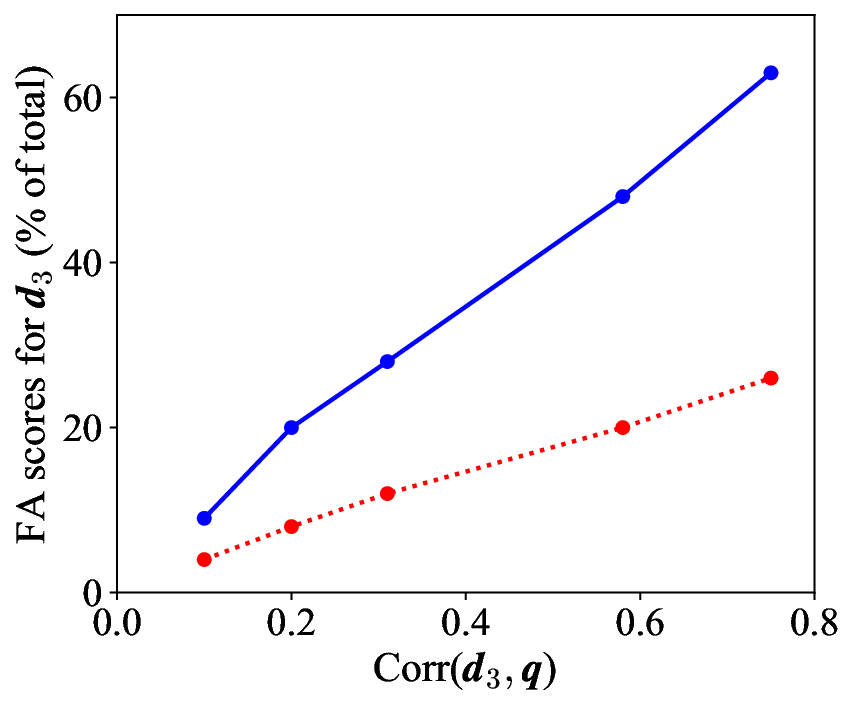}
\caption{\textbf{Variation of the contribution fraction with correlation.} The relative contribution of the Sensor 3 inputs, $\boldsymbol{d}_3$, to the total feature attribution (FA) magnitude is compared for Naive FA (red dashed) and Clustered FA (blue solid) as the correlation between $\boldsymbol{d}_3$ and the target $\boldsymbol{q}$ increases.}
\label{fig:attr_demo}
\end{figure}

As the input-output correlation coefficient ``Corr($\boldsymbol{d}_i$, $\boldsymbol{q}$)'' of a cluster center increases, with other sensors decreasing to maintain total information, both the Naive and Clustered FA for that sensor scale near-linearly with the correlation (Fig.~\ref{fig:attr_demo}). However, the ``Clustered FA'' score exhibits a steeper gradient, indicating a shift in relative importance among the remaining sensors when the correlated sensor is removed. Namely, the difference between ``Naive FA'' and ``Clustered FA'' increases as ``Corr($\boldsymbol{d}_i$, $\boldsymbol{q}$)'' increases, demonstrating that grouping correlated candidates and evaluating only the cluster centers amplifies the distinction between influential and redundant sensors. This effect arises because when naive FA is applied, duplicated information among correlated features dilutes the apparent importance of the uncorrelated ones. Consequently, incorporating clustering widens the FA gap between critical and non-critical candidates, enhancing the discriminative power of the selection process.

This behavior underscores a key advantage of CAAF. By condensing correlated sensors into representative cluster centers, the framework magnifies the relative importance of truly informative features, making optimal sensor selection more robust and reliable.

\begin{table}[t]
    \centering
    \caption{\textbf{Statistics for the synthetic data experiment.}}
    \begin{tabular}{|l|c|c|c|}
        \hline
        Sensor Index & 1 & 2 & 3 \\ 
        \hline
        Corr($\boldsymbol{d}_i$, $\boldsymbol{d}_1$) & 1.0 & 0.9 & 0.0 \\ 
        \hline
        Corr($\boldsymbol{d}_i$, $\boldsymbol{q}$) & 0.65 & 0.89 & 0.32  \\ 
        \hline
        Naive FA & 29\% & 59\% & 12\% \\ 
        \hline
        Cluster Label & 0 & \textbf{0} & \textbf{1}  \\ 
        \hline
        Clustered FA & & 76\% & 24\%  \\ 
        \hline
    \end{tabular}
    \caption*{{\footnotesize``Corr($\boldsymbol{d}_i$, $\boldsymbol{d}_1$)'' shows the correlation of each sensor's inputs ($\boldsymbol{d}_i$) with those of Sensor 1 ($\boldsymbol{d}_1$). ``Corr($\boldsymbol{d}_i$, $\boldsymbol{q}$)'' shows the level of correlation between the sensor inputs and the target variable ($\boldsymbol{q}$). ``Naive FA'' indicates the feature attribution (FA) percentages (or importance scores) of the individual candidate sensor inputs by directly performing FA on a model trained on all candidate features without the preceding clustering. ``Cluster Label'' reveals the clusters formed using the Affinity Propagation clustering algorithm. The boldfaced labels indicate the cluster centers. ``Clustered FA'' prints the CAAF results, which are the FA percentages of only the cluster centers.}}
    \label{tab:21_demo}
\end{table}

\subsection*{Structural health monitoring on a cantilever beam}
Having illustrated the working principles of CAAF, we now evaluate its performance on practical sensor placement applications. This section demonstrates CAAF’s applicability in identifying the most informative elements on a cantilever beam for SHM. SHM ensures the safety and longevity of structures (bridges, buildings, etc.) by detecting damage and optimizing maintenance through sensor signals. Determining the best locations for a restricted number of sensors within a structure is a key step in acquiring precise, real-time measurements. Specifically, in structural analysis, sensor positions must be carefully selected so that the authentic mode shapes and target frequencies can be accurately recovered \cite{papadopoulos1998sensor}. Additionally, SHM strategies rely on OSP techniques for various applications, including finite-element updating, state estimation, and damage localization \cite{friswell2015,liu2018optimal}. Nonetheless, identifying OSP for SHM presents significant difficulties because it requires maximizing information gain about all potential states across the entire structure, while minimizing costs associated with the sensor equipment and computation.

\begin{figure}[t]
\centering
\includegraphics[width=1.0\textwidth]{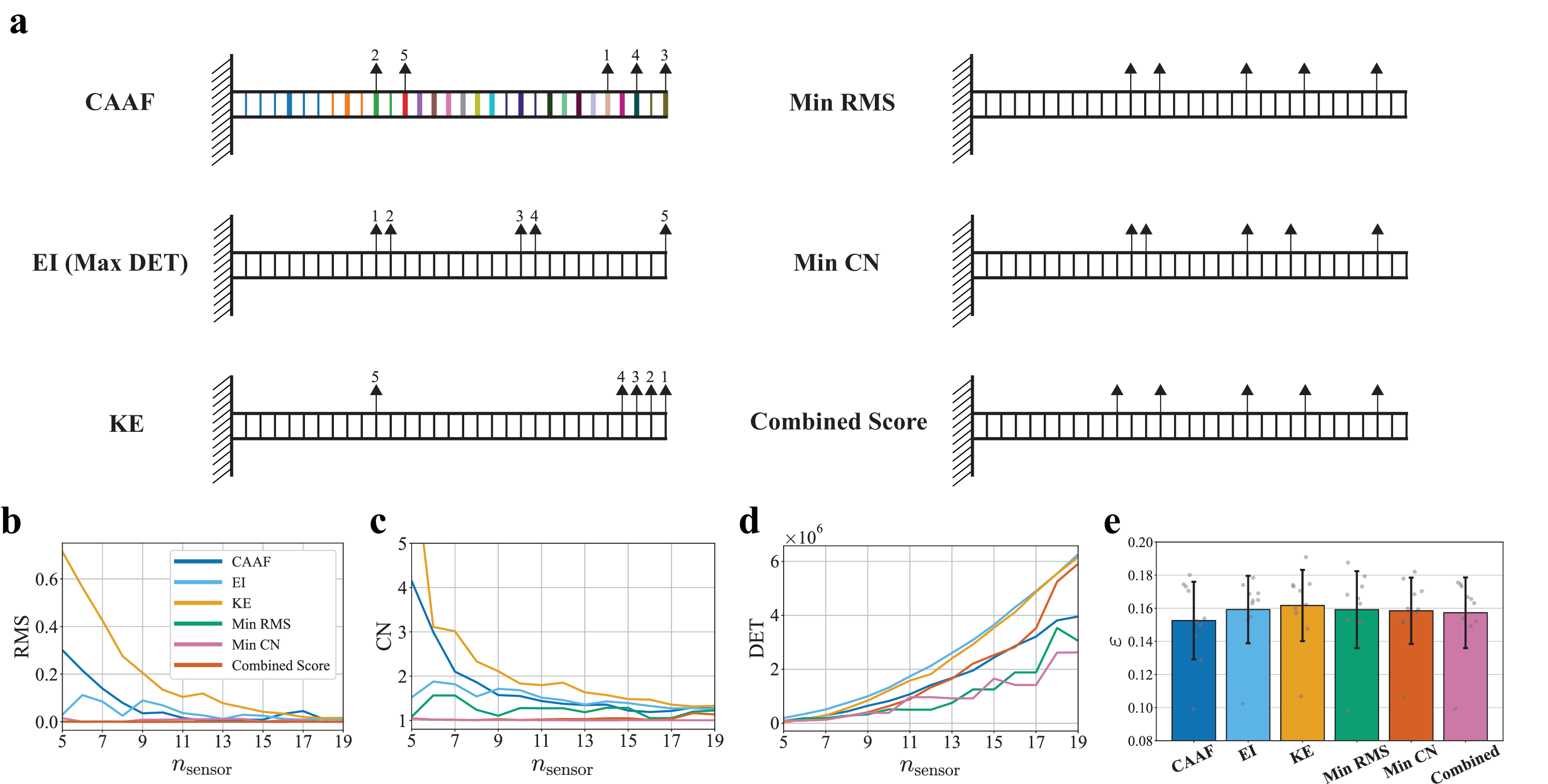}
\caption{\textbf{Locations and performance of structural health monitoring (SHM) sensors.} 
\textbf{a} Five SHM sensors on the cantilever beam identified via CAAF, Effective Independence (EI) or Maximum Determinant (Max DET), Kinetic Energy (KE), and brute-force methods, including Minimum Root Mean Square (Min RMS), Minimum Condition Number (Min CN), and Combined Score. For the CAAF method, the locations of 19 node clusters (colored bars) and their respective centers (thick colored bars) are marked. Numerical labels indicate the importance rankings for methods capable of providing a sequential ranking. 
\textbf{b--d} Performance variation is measured using (\textbf{b}) RMS, (\textbf{c}) CN, and (\textbf{d}) DET versus the number of sensors ($n_\text{sensor}$) for configurations determined by CAAF, EI, KE, Min RMS, Min CN, and Combined Score. 
\textbf{e} Comparison of the mean relative $L_2$ error and associated standard deviations for modal coefficient prediction across different sensor identification methods, evaluated over $n=10$ training runs.}
\label{fig:SHM_combined}
\end{figure}

We investigate the sparse placement of sensors on a uniform cantilever beam, a classic benchmark scenario for SHM. By applying CAAF, we select 5 optimal sensor locations from 30 candidate nodes on the cantilever beam, considering only the first three mode shapes and translational degrees of freedom. The CAAF uses the Affinity Propagation (AP) algorithm to perform clustering and identify cluster centers. As shown in Fig.~\ref{fig:SHM_combined}a, AP groups the candidate sensor nodes into 19 clusters and captures the correlations among nodes located near the fixed joint. Next, the CAAF employs a multilayer perceptron (MLP) model trained to predict the modal coefficients of the first three modes from nodal deflection measurements at the candidate sensor locations. 

OSP methods for SHM have been extensively studied, with common approaches including EI \cite{kammer1991sensor} and its variants (EI-mass \cite{garvey1996evaluation}, EI-dpr \cite{meo2005optimal}), as well as energy-based methods like kinetic energy (KE) and weighted average KE \cite{kammer1991sensor,li2007connection,liu2018optimal}. In this study, we validate the data-driven CAAF method against these established analytical OSP approaches using the standard SHM assessment metrics on the cantilever benchmark. The three traditional assessment metrics are root mean square (RMS), condition number (CN), and determinant (DET). They measure the RMS of the off-diagonal entries of the mass-weighted modal assurance criterion (MMAC) matrix, CN of the mode shape matrix, $\boldsymbol{\Phi}$, and DET of the reduced Fisher information matrix, respectively. RMS quantifies the orthogonality of the mode shape matrix, truncated to the selected sensor locations, while the CN and DET assess modal independence. Optimal sensor configurations should minimize RMS and CN while maximizing DET, thereby ensuring maximal mode shape orthogonality and independence. High RMS/CN values or low DET values indicate suboptimal sensor placement. To compare the sensors from OSP methods with the optimal ones, we also identified sensor configurations that achieve the optimum for each of these three metrics using a brute-force search by minimizing RMS (Min RMS), CN (Min CN), and the Combined Score, defined as
\begin{gather}
    \text{Combined Score} = \frac{RMS}{RMS_{\text{ref}}} + \frac{CN - 1}{CN_{\text{range}}} + \left( 1 - \frac{DET}{DET_{\text{max}}} \right),
\label{combined_score}
\end{gather}
where the reference RMS ($RMS_{\text{ref}}$), the reference CN range ($CN_{\text{range}}$), and the maximum determinant ($DET_{\text{max}}$) are pre-computed normalization constants. Since the sensors that maximize the DET are mathematically equivalent to those identified by the EI method, these are referred to collectively as EI throughout this application. The sensor configurations identified by EI, KE, and the brute-force methods are displayed in Fig.~\ref{fig:SHM_combined}a. The \textit{Methods} section provides complete implementation details for these metrics. 

Sensor configurations derived from CAAF are compared against alternative methods using the three established metrics. As illustrated in Fig.~\ref{fig:SHM_combined}b--d, CAAF achieves performance comparable to the EI and KE methods in both RMS and CN for $n_\text{sensor}>5$. Although the DET values for CAAF-derived configurations are lower at higher sensor densities, they exhibit strong agreement with benchmark methods in sparse sensing regimes. This performance pattern confirms that CAAF-based placements preserve mode shape orthogonality and independence while maintaining competitive information content. We also observe that brute-force sensors (Min RMS, Min CN, and Combined Score) achieve very low RMS and CN at the expense of DET performance, suggesting an inherent trade-off between these metrics. To quantify the predictive performance in these applications, we utilize the relative $L_2$ prediction error for the target quantities of interest ($\boldsymbol{q}$), denoted as $\varepsilon$ and defined as:
\begin{gather}
    \varepsilon=\frac{\big\| \widehat{\boldsymbol{q}}-\boldsymbol{q} \big\|_2}{\big\|\boldsymbol{q}-\langle\boldsymbol{q}\rangle\big\|_2}~,
    \label{equ:L2_error}
\end{gather}
where $\widehat{~\cdot~}$ represents the predicted values, $\langle\cdot\rangle$ indicates sample-averaged values, and $\big\|\cdot \big\|_2$ denotes the $L_2$ norm. Notably, while CAAF sensors show slightly lower values in analytical metrics, they demonstrate superior prediction accuracy when integrated into ML-based modeling for modal coefficients, as evidenced by the minimum prediction error shown in Fig.~\ref{fig:SHM_combined}e.

\subsection*{Lift prediction for airfoils under gusts}
In active flow control applications, particularly gust mitigation for small aircraft and unmanned aerial vehicles, controller performance largely depends on accurate flow sensing. Turbulent winds and complex terrain generate unpredictable aerodynamic disturbances, exacerbated by the small size and low inertia of these vehicles \cite{boettcher2003, granlund2014,galway2008}. Effective sensing and control strategies are essential to maintain stability, but they face significant challenges due to the stringent requirements for accurate perception and high-bandwidth, low-latency response in highly unsteady realistic turbulence. OSP becomes critical in this context, as it directly affects the observability of key flow features and the robustness of the control performance. Here, we focus on identifying OSP on an airfoil for surface pressure sensors to enable precise temporal predictions of the airfoil lift coefficient ($C_L$), which is crucial for mitigating gust-induced lift fluctuations \cite{renn2022,beckers2024deep,nair2023,paris2021robust}.

Using incompressible large-eddy simulations (LES), we simulated flow over a NACA 0012 airfoil under both gusty and non-gusty conditions, collecting time-resolved airfoil surface pressure distributions and lift coefficients. Pressure data from the LES are averaged over the spanwise direction, as only the two-dimensional placement is investigated. The CAAF is then applied to determine optimal sensor locations, which involves clustering the candidates and training an MLP model with the LES data to estimate the instantaneous airfoil lift coefficient using pressure measurements at the candidate sensor locations. The predictive accuracy of the CAAF sensor configuration is benchmarked against those derived from naive FA, Pivoted-QR, Bayesian experimental design, and uniform sensor distribution methods.

Following our previous work \cite{leung2024}, gusty inflow conditions are generated by placing a cylinder directly upstream of the airfoil. The Reynolds number $Re_c=U_\infty C/\nu$, where $C$ denotes the chord length, $U_\infty$ is the freestream velocity, and $\nu$ represents the kinematic viscosity, is set to $10^4$. We simulate five distinct flow configurations, systematically varying cylinder geometry and angle of attack (AoA): (1) ``None-5'' (baseline, AoA = 5\degree), (2) ``None-11'' (baseline, AoA = 11\degree), (3) ``Cylinder-5'' (with cylinder, AoA = 5\degree), (4) ``Cylinder-11'' (with cylinder, AoA = 11\degree), and (5) ``Cylinder2-0'' (modified cylinder geometry, AoA = 0\degree). The complete geometric configuration is detailed in the \textit{Methods} section. For our analysis, we use the ``None-5'', ``None-11'', and ``Cylinder-5'' cases for model training and sensor identification, reserving the ``Cylinder-11'' and ``Cylinder2-0'' cases exclusively for testing purposes.

\begin{figure}[t]
    \centering
    \makebox[\textwidth][c]{\includegraphics[width=1\textwidth]{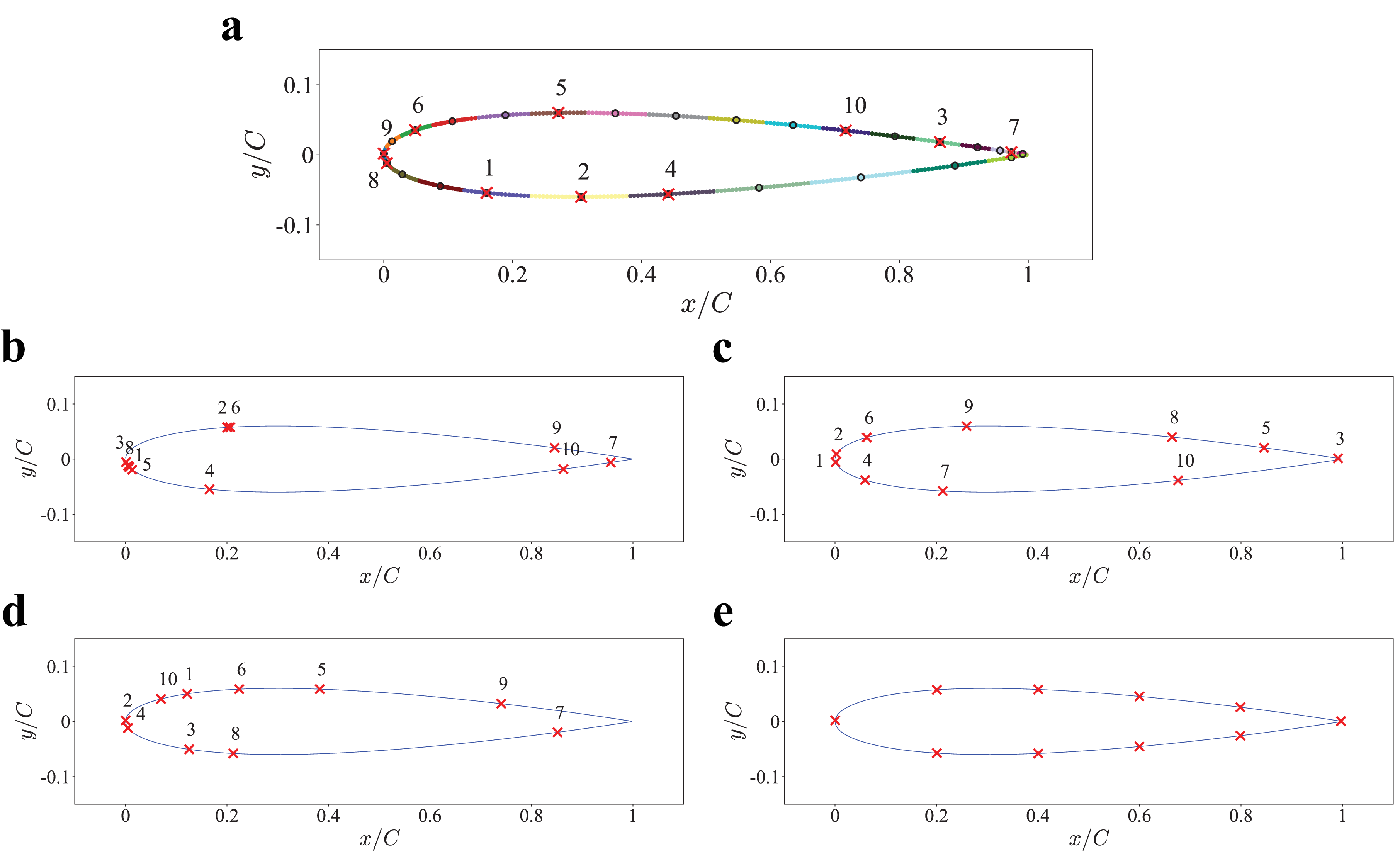}}
    
    \caption{\textbf{Airfoil surface pressure sensor locations.} 
    \textbf{a} Optimal 10-sensor configuration identified by CAAF, including 27 identified clusters (colored dots) and their respective centers (black circles).
    \textbf{b--e} Comparative sensor configurations identified via (\textbf{b}) naive FA, (\textbf{c}) Pivoted-QR, (\textbf{d}) Bayesian experimental design, and (\textbf{e}) uniform distribution. 
    In all panels, red crosses mark the locations of the sensors. 
    Numerical labels indicate importance rankings corresponding to selection order; labels are omitted for (\textbf{e}) uniform distribution and do not imply selection order for the non-sequential (\textbf{c}) Pivoted-QR.}
    \label{fig:airfoil_sensors}
\end{figure}

\begin{figure}[t]
\centering
\makebox[\textwidth][c]{\includegraphics[width=1\textwidth]{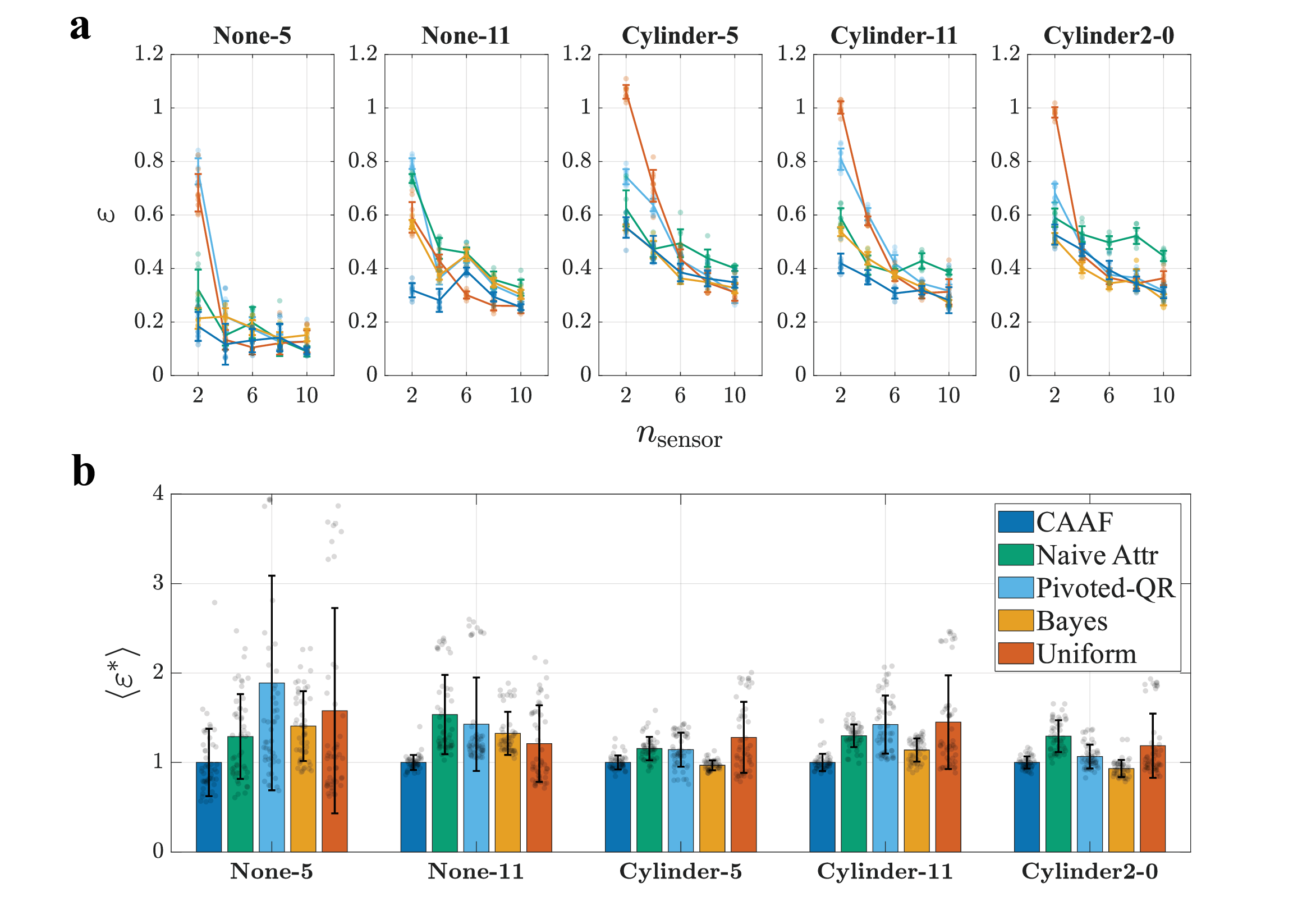}}
\caption{\textbf{Prediction performance comparison between sensor configurations identified using different OSP methods.} 
\textbf{a} Lift prediction error across various sensor counts ($n_{\text{sensor}}$) using locations identified via CAAF, naive FA (Naive Attr), POD-based QR-pivoting (Pivoted-QR), Bayesian experimental design (Bayes), and a uniform distribution (Uniform). Color mapping corresponds to the legend in \textbf{b}.
\textbf{b} Normalized lift prediction error averaged over all sensor counts, $\left\langle\varepsilon^* \right\rangle$. Error bars denote one standard deviation, centered on the mean value of the plotted quantity computed from at least $n=10$ independent training runs with different random seeds.}
\label{fig:error_sensors}
\end{figure}

We leverage the AP algorithm again to implement clustering for CAAF, with results visualized in Fig.~\ref{fig:airfoil_sensors}a. It is observed that spatial correlation among pressure measurements leads to clustering based on proximity, with higher cluster density near the leading and trailing edges due to greater pressure variability.

The optimal 10-sensor configurations identified by the different OSP methods are displayed in Fig.~\ref{fig:airfoil_sensors}. Notably, the configurations derived from data-driven approaches exhibit distinct spatial distributions, reflecting their different underlying selection criteria. Nonetheless, some common features of the CAAF configurations can be observed in the others and explained by analyzing the physics of the flow around an airfoil. In addition to the large separation bubble that typically forms on the suction surface near the trailing edge of the airfoil, where most fluctuations occur due to re-circulation, informative regions also include the stagnation point just below the leading edge corresponding to the location of the first sensor on the CAAF configuration in Fig.~\ref{fig:airfoil_sensors}a. The pressure around the stagnation point varies as the effective AoA deviates from the true AoA of the airfoil under the cylinder wake. Since the effective AoA is highly correlated with lift fluctuations, such an area is expected to play a key role in lift inference. This may explain the strong performance of the CAAF sensors in extremely sparse configurations, as the top-ranked CAAF sensors are positioned near the stagnation region. This is particularly expected in the high-AoA cases of ``None-11'' and ``Cylinder-11'', where the stagnation point shifts downstream toward these sensor locations leading to a more informative pressure surface. Unique to the cases with gusty inflows, separation bubbles appear around the leading edge of the airfoil (see Supplementary Fig.~S1 for flow field visualization). The flow characteristics within the separation region are significantly different, and the separation bubble size greatly impacts the airfoil lift coefficient. Considering this, the CAAF design recognizes it by prioritizing leading-edge sensor placement to capture these flow dynamics.



We evaluated the predictive performance of each optimal sensor configuration by training dedicated MLP models using only pressure measurements at the selected sensor locations. The prediction accuracy of the trained models acts as a direct metric for evaluating the optimality of each sensor configuration. Figure~\ref{fig:error_sensors}a shows the variation of the prediction error $\varepsilon$ for $C_L$ with increasing number of sensors. The results demonstrate that CAAF-generated sensors outperform other methods in sparse placement regimes across all five geometries tested, while achieving comparable performance at higher sensor counts where different approaches converge. Overall, the CAAF-derived sensors exhibit robust performance across a wide range of flow scenarios and sensor counts, as evidenced by the small error variation. Figure~\ref{fig:error_sensors}b highlights that CAAF achieves the highest prediction accuracy for the ``None-5'', ``None-11'', and ``Cylinder-11'' cases, while only marginally trailing Bayesian experimental design in the remaining configurations. Collectively, these results establish CAAF as a consistent framework for maximizing predictive performance compared to other benchmarks. This superior performance confirms the effectiveness of CAAF for OSP in airfoil lift prediction tasks and suggests its broader applicability to similar target inference problems.

\subsection*{Wall-normal velocity estimation in turbulent flows}
Precise knowledge of the off-wall wall-normal velocity ($v$) in wall-bounded turbulence is highly valuable, as it enables active control of boundary layer flow structures via fluidic actuation to achieve drag reduction, potentially yielding substantial energy savings in various engineering applications \cite{choi1994}. Since making accurate non-intrusive measurements of the off-wall velocities is impractical, previous efforts \cite{guastoni2021convolutional, park2020machine, guemes2019sensing, milano2002neural} have focused on predicting the off-wall velocities using wall quantities such as the instantaneous streamwise and spanwise velocity wall-normal gradients at the wall ($\left.\partial u / \partial y\right|_w$ and $\left.\partial w / \partial y\right|_w$), and the wall pressure ($p_w$). However, prediction remains challenging due to the inherently nonlinear and chaotic nature of turbulent flows, particularly when relying on sparse measurements as required in real-world deployments, where most existing methods assume access to full-field inputs. Therefore, the objective of this application is to identify optimal wall sensing locations where wall pressure measurements provide the most informative input for estimating off-wall wall-normal velocity in turbulent channel flows.

\begin{figure}[t]
    \centering
    \makebox[\textwidth][c]{\includegraphics[width=1\textwidth]{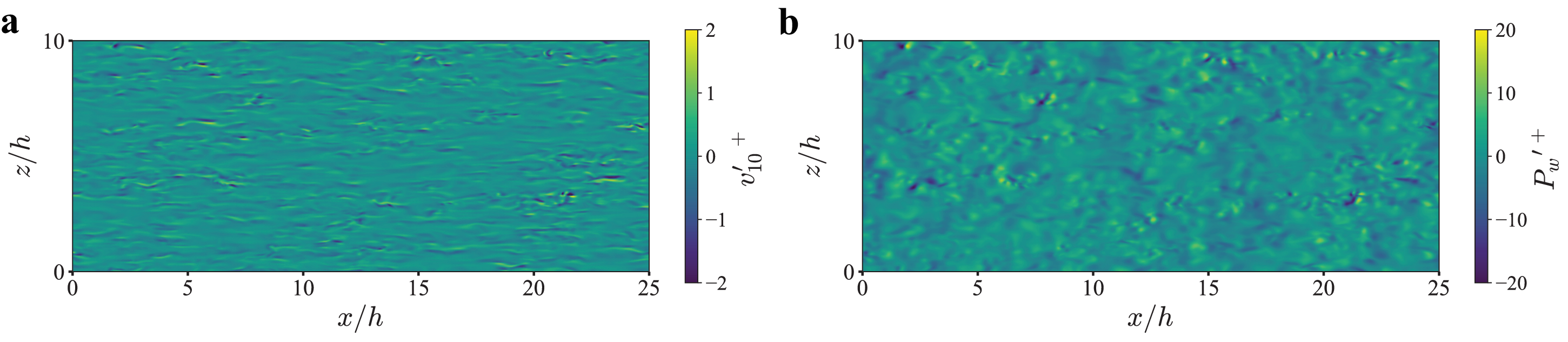}}

\caption{\textbf{Instantaneous data visualization from DNS of channel flow.} Non-dimensionalized (\textbf{a}) wall-normal velocity fluctuation at $y^+=10$, $v_{10}'^+$, and (\textbf{b}) wall pressure fluctuation, $p_w'^+$.}
\label{fig:PV_fields}
\end{figure}

Following a previous study \cite{park2020machine}, candidate locations of wall pressure probes for wall-normal velocity estimation are sampled from streamwise and spanwise positions within $-50<\Delta x^+,\Delta z^+ < 50$, where the superscript $+$ denotes that the quantity is nondimensionalized by viscous units defined based on the kinematic viscosity and friction velocity. We employed exclusively the wall pressure as input, as it shows a high correlation with off-wall wall-normal velocity in the previous study \cite{park2020machine}. Instantaneous wall-normal velocity at the wall-normal location $y^+=10$ and center of the sensing plane ($\Delta x^+,\Delta z^+=0$), denoted by $v_{10}$, is predicted by a temporal convolutional network (TCN) \cite{lea2016temporal} model using the wall pressure measurements (at $y^+=0$) as inputs at each time snapshot. The location $y^+ = 10$ is selected based on prior studies \cite{choi1994} indicating it as the optimal height for implementing effective opposition control for drag reduction under the tested conditions. The reference data is obtained through a direct numerical simulation (DNS) of an incompressible turbulent channel flow at friction Reynolds number $Re_\tau \approx 186$. Figure~\ref{fig:PV_fields} depicts the instantaneous fields of the wall pressure and wall-normal velocity fluctuation at $y^+ = 10$ obtained from the DNS. The highly nonlinear correlation between these quantities makes predicting wall-normal velocity a particularly nontrivial task.

\begin{figure}[t]
    \centering
    \makebox[\textwidth][c]{\includegraphics[width=1.0\textwidth]{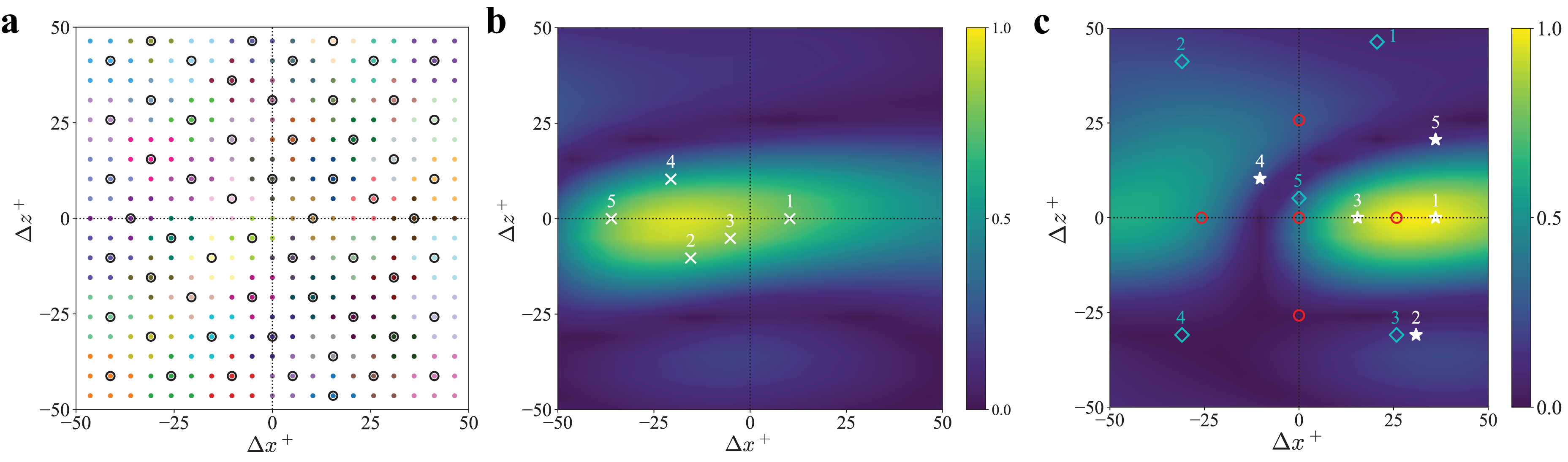}}

\caption{\textbf{Clustering and optimal sensor configuration for wall-normal velocity estimation.} 
\textbf{a} Initial candidate sensor locations (361 probes) and AP clustering results (reduced to 58 representative locations). Cluster members are shown as colored dots with cluster centers indicated by black open circles. 
\textbf{b} Normalized cross-correlation between $p_w$ and $v_{10}$ at a time lag of $\Delta t_{lag}^+ \approx -4.3$, overlaid with the top five CAAF sensors (white crosses).
\textbf{c} Zero-lagged $p_w-v_{10}$ cross-correlation overlaid with five Bayesian experimental design sensors (white stars), five Pivoted-QR sensors (cyan diamonds) and the uniform five-sensor benchmark (red circles).
The numerical labels indicate the importance rankings of the data-driven sensors.}

\label{fig:PV_sensors}
\end{figure}

As shown in Fig.~\ref{fig:PV_sensors}a, the clustering analysis condenses the initial 361 probe locations to 58 representative candidates through AP. The clusters exhibit localized circular shapes consistent with the wall pressure spatial autocorrelation patterns reported in the literature \cite{choi1990space, liu2024wall}. By recognizing these features inherent to the problem, the clustering algorithm naturally groups points according to the underlying physical patterns of the flow, providing meaningful data representatives. 

With the established cluster representatives and the trained TCN model for predicting $v_{10}$, CAAF is successfully applied. The optimal five-sensor configuration is shown in Fig.~\ref{fig:PV_sensors}b. To illustrate the physical basis for this sensor selection pattern, we present contours of the two-point cross-correlation coefficient between $p_w$ and $v_{10}$, defined as
\begin{gather}
    C_{v p}\left(\Delta x, \Delta z, \Delta t_{lag}\right)=\frac{\left\langle v_{10}'\left(x, z, t\right) p_w'\left(x+\Delta x, z+\Delta z, t+\Delta t_{lag}\right)\right\rangle}{\sigma_{v_{10}'}(x,z)\sigma_{p_w'}(x+\Delta x,z+\Delta z)},
\end{gather}
where $(\cdot)'$ denotes fluctuations, $\langle \cdot \rangle$ represents averaging over temporal and homogeneous directions, $\sigma$ denotes the standard deviation, and $\Delta t_{lag}$ indicates the time lag applied to the wall pressure. As shown in Fig.~\ref{fig:PV_sensors}c, the zero-lagged $v_{10}$-$p_w$ cross-correlation reveals a distinct high-correlation region downstream of the prediction origin, which is largely symmetric about the $\Delta z^+=0$ axis. A secondary, slightly weaker high-correlation region appears upstream, centered around $\Delta x^+ \approx -25$ to $-50$. Similar spatial correlation structures have been reported in channel flows at $Re_\tau = 178$ and $578$ by Park and Choi \cite{park2020machine}, who also visualized a saliency map from a convolutional neural network trained to predict $v_{10}$ from $p_w$. The CAAF sensor locations align closely with the cross-correlation contour at a time lag of $\Delta t_{lag}^+ \approx -4.3$ (Fig.~\ref{fig:PV_sensors}b), where the wall pressure precedes the velocity. Under this time lag, the high-correlation region shifts upstream of the prediction target. This alignment suggests that optimal sensing sites should be positioned upstream to account for the convective behavior of the flow. Furthermore, since the TCN leverages time-history information, providing sufficient look-ahead data is essential for accurate prediction. Prior literature on opposition control \cite{luhar2014opposition, lee2015opposition} has similarly observed enhanced wall-normal velocity control performance using time-delay or phase-shift techniques. Specifically, Lee \cite{lee2015opposition} reported improved control performance when sensors were placed at approximately $\Delta x^+ = -25$ , which validates the characteristic time lag observed in our results.
The CAAF-identified sensors exhibited physically interpretable placement, aligning with these flow features and supporting their relevance for wall-normal velocity prediction, as indicated by the cross-correlation statistics and prior literature.


As a benchmark configuration, we consider an array of five sensors uniformly positioned within the sensing plane, highlighted in Fig.~\ref{fig:PV_sensors}c. For a comprehensive evaluation, optimal five-sensor configurations were also identified using Bayesian experimental design (Fig.~\ref{fig:PV_sensors}c) and Pivoted-QR (Fig.~\ref{fig:PV_sensors}c). The Bayesian sensor locations exhibit strong spatial agreement with the high-correlation region of the zero-lagged $v_{10}$-$p_w$ cross-correlation, thereby validating the method. The average predictive performance for the wall-normal velocity time series, comparing CAAF-derived, Bayesian, Pivoted-QR, and uniform configurations, is summarized in Table~\ref{tab:PV_table}. The CAAF sensors achieve a $L_2$ prediction error of 0.547 and a correlation of 0.837 with the reference, outperforming the benchmark arrays. Supplementary Fig.~S2 contrasts the predicted wall-normal velocity sequences for a selected section of the testing dataset. Although tracking the reference signal with only five sensors is challenging, the CAAF-identified sensors demonstrate more consistent tracking of the wall-normal velocity than the benchmark configurations, particularly during periods of large fluctuations.

These results demonstrate that CAAF effectively identifies optimal sensor locations for wall-normal velocity estimation. More specifically, the CAAF-derived sensor configuration outperforms the alternative benchmarks and achieves similar predictive accuracy to the full-field learning approach reported in the literature \cite{park2020machine}.

\begin{table}[t]
    \centering
    \caption{\textbf{Wall-normal velocity prediction performance comparison between different sensor configurations.}}
    \begin{tabular}{|l|c|c|c|c|c|}
        \hline
        Sensor configuration & cluster centers & CAAF & Pivoted-QR & Bayes & uniform  \\ 
        \hline
        Corr($v_{10}$, $\widehat{v_{10}}$) &  0.953 & 0.837 & 0.765 & 0.806 & 0.800 \\ 
        \hline
        $L_2$ error & 0.302 & 0.547 & 0.644 & 0.592 & 0.600 \\ 
        \hline
    \end{tabular}
    \caption*{{\footnotesize Performance metrics for configurations using 58 cluster centers, 5 CAAF sensors, 5 POD-based QR-pivoting (Pivoted-QR) sensors, 5 Bayesian experimental design (Bayes) sensors, and 5 uniform sensors. Corr($v_{10}$, $\widehat{v_{10}}$) indicates the average correlation coefficient between referenced and predicted wall-normal velocity. $L_2$ error denotes the average normalized $L_2$ error as defined in Eq.~\ref{equ:L2_error} applied to $v_{10}$. Both metrics are averaged over $n=5$ training runs with different random seeds.}}
    \label{tab:PV_table}
\end{table}

\section*{Discussion}
The validation cases presented in the image clustering and synthetic data experiments illustrate that the additional clustering step not only reduces the problem's dimensionality to improve computational efficiency but also enhances the effectiveness of FA for sensor placement in correlated datasets. Direct FA on the original candidate set tends to identify the global attribution maximum and its nearest neighbors, which often exhibit high correlation between sensor inputs and thus lead to redundant sensor selections. In contrast, by computing FA on cluster centers, the CAAF method prioritizes candidates with locally maximal FA values, collectively enabling a more comprehensive measurement strategy.

It is also worth noting that CAAF assumes no inherent restrictions on the positioning of the sensors. In addition, practical limitations, such as constraints imposed by geometry or the infeasibility of placing sensors at certain locations, can be readily incorporated into the sensor identification process. This can be achieved by excluding candidate sensor locations that are deemed infeasible before applying the framework; for example, sensors at the trailing edge of the airfoil are often omitted due to sharpness and can thus be removed prior to the clustering step.
The framework demonstrates broad applicability and competitive performance for application-oriented sensor placement problems. We showcase its effectiveness in predicting different quantities of interest, ranging from modal coefficients in solid structures to aerodynamic coefficients and velocity components in wall-bounded turbulent flows. Sensor configurations selected by CAAF yield performance comparable to and often superior to those obtained using analytical or empirical methods. In contrast to these alternative techniques, CAAF demonstrates distinct advantages by yielding robust sensor placements that are less sensitive to prior assumptions and more effective in capturing complex nonlinearities than Bayesian experimental design. Furthermore, by explicitly incorporating the prediction target into its formulation, CAAF ensures superior predictive performance compared to target-agnostic approaches such as Pivoted-QR. As a purely data-driven approach, the framework is highly adaptable across diverse domains, owing to its decoupled clustering, model development, and FA stages. 

However, this flexibility also introduces potential limitations.
The quality of the resulting sensor configurations depends critically on the clustering and modeling processes, which may require careful tuning and domain expertise due to the sensitivity of the final configuration to the density and quality of the cluster centers. Optimal performance necessitates clustering strategies informed by physical insights and domain-specific correlation metrics that reflect dynamical interdependencies. Specifically, cluster centers must sufficiently populate informative regions, such as the leading and trailing edges of an airfoil, while maintaining a total cluster count and spacing that avoids overly sparse configurations or excessive feature dimensionality, which may complicate training or induce redundant sensors. In the aforementioned airfoil lift and wall-normal velocity applications, the clustering results align with physical intuition derived from airfoil theory or correlation analysis, demonstrating their interpretability through domain-specific logic. Therefore, the choice of clustering and FA algorithms must also be carefully considered to ensure robust and interpretable outcomes.
Furthermore, as CAAF identifies optimal sensor locations via the learned mapping of a converged model, the framework’s efficacy relies on the accuracy of the surrogate model in capturing input-output relationships, which is influenced by data quality, model architecture, and hyperparameter selection. To ensure maximum robustness, it is recommended to train the model over multiple runs with varying random seeds and data shuffling, then compute the ensemble-averaged attribution scores. As the objective of CAAF is to optimize sensor placement for predictive performance, alterations in the prediction method or model formulation are likely to influence the resulting sensor configurations. While the architecture and training process used in this study may not be fully optimized, the models still achieve sufficient accuracy to generate sensor configurations with satisfactory performance. Further optimization of the modeling approach could yield even higher-performing sensor configurations. 
This study demonstrates CAAF primarily for identifying OSP in the prediction of low-dimensional quantities of interest based on high-fidelity simulation datasets. Its generalizability is partially demonstrated in the airfoil lift prediction application, where sensors identified on a subset of cases are successfully evaluated under unseen conditions. While this framework is theoretically applicable to noisy experimental observations and high-dimensional or autoregressive targets, its validation remains limited to the cases presented here. These potential extensions represent new directions for future investigations.

Due to the rigorous requirement for model accuracy, the CAAF framework can be computationally intensive. For a dataset of 50,000 input-output pairs, the complete CAAF pipeline requires approximately ten minutes using a small-scale MLP model trained on an RTX 5070 GPU. For the same data volume, the Bayesian experimental design method completes within several minutes utilizing GPU-optimized operations. In contrast, Pivoted-QR represents the most computationally efficient method, with execution times on the order of seconds.

Although CAAF offers no direct computational cost benefits, its utility stems from the use of ML-based models and its data-driven approach. The ML-based framework allows researchers to leverage pre-existing inference pipelines or pre-trained models. Additionally, its purely data-driven methodology means that the sensor identification process requires no further experiments or system interactions once the initial data is collected. Despite its limitations, CAAF effectively determines sensing strategies for prediction tasks essential to the control of diverse systems. The overall performance of the CAAF-derived sensors exhibits better accuracy in quantity inference tasks in the tested applications. The data-driven nature of CAAF enables its application to a wide range of scenarios without requiring major modifications. By simply adjusting the set of candidate sensor locations, we can identify OSP in three-dimensional configurations and other more complex geometries beyond those examined in this study.

\section*{Methods}

\subsection*{Affinity Propagation}
AP is a clustering algorithm that identifies exemplars (representative data points) through iterative message passing between pairs. It utilizes two types of messages: responsibilities $r(i,k)$, which indicate how suitable point $k$ is as an exemplar for point $i$ compared to other candidates, and availabilities $a(i,k)$, reflecting how appropriate it is for $i$ to choose $k$ as its exemplar based on support from other points. The algorithm initializes availabilities to zero and iteratively updates responsibilities and availabilities until convergence, as described carefully in the original paper \cite{frey2007clustering}. Exemplars emerge as points where $a(i,i)+r(i,i)$ is maximized. The algorithm requires no preset cluster count and handles non-metric or asymmetric similarities, efficiently converging to high-quality clusters. Since AP inherently identifies cluster centers through exemplars, it is particularly well-suited for the CAAF, which relies explicitly on the use of cluster centers. In contrast, when alternative clustering algorithms are employed, cluster centers can be determined through other methods, such as selecting the candidate with the highest average correlation to all other members within each cluster.

\subsection*{Integrated Gradients}
For our task of finding optimal sensor locations, we chose to use the IG algorithm due to its two desirable properties: sensitivity and implementation invariance \cite{sundararajan2017}. Sensitivity ensures that a feature's impact on the model's output is proportionally conveyed by the FA value; the implementation invariance provides robustness to variations in the model's architecture and implementation. The IG algorithm aggregates the model's gradients with respect to the input features along a linear path connecting a predefined baseline (often chosen to be the zero vector) with the input data, \emph{i.e.},
\begin{gather}
    \text{IG} (d_i) := \int_{0}^1 \frac{\partial F(\alpha \boldsymbol{d})}{\partial \alpha d_i} d_i \mathrm{d} \alpha~,
\end{gather}
where $\alpha \in \left[ 0,1 \right]$ is a scaling parameter that interpolates between the baseline and the input, $\boldsymbol{d}$ represents the data point containing the inputs for all sensors $i=1, \dots,M$, where $M$ is the number of candidate sensors, and $F(\cdot)$ is a data-driven model such that $q \approx F(\boldsymbol{d})$. In our implementation, we adopt the zero vector as the baseline. The magnitude of the FA scores, averaged over all data points, is computed for each candidate sensor and subsequently ranked to identify the most contributing sensors. While other FA algorithms, such as GradientSHAP and KernelSHAP, were also tested, they did not provide significant advantages over the IG method in terms of the performance of the resultant sensor configuration.

\subsection*{Pivoted-QR}
For the Pivoted-QR method, a singular value decomposition is performed on the data matrix $ \boldsymbol{d}\in \mathbb{R}^{N_t \times M}$, defined as
\begin{gather}
    \label{equ:data_matrix}
    \boldsymbol{d} = \begin{bmatrix}
     d_1^{\left ( 1 \right )}&  d_2^{\left ( 1 \right )} & \cdots & d_{M}^{\left ( 1 \right )}\\ 
    \vdots &  \vdots  & &\vdots \\ 
    d_1^{\left ( N_t \right )}&  d_2^{\left ( N_t \right )} & \cdots & d_{M}^{\left ( N_t \right )}
    \end{bmatrix},
\end{gather}
where $d_i^{(j)}$ is the pressure measurement given by the $i$-th sensor at the $j$-th snapshot, with $N_t$ total time snapshots. The decomposition yields the predominant modes in the data as illustrated by
\begin{equation}
\boldsymbol{d} = \boldsymbol{\Psi}\boldsymbol{\Sigma} \boldsymbol{\Phi}^T, 
\end{equation}
where $\boldsymbol{\Psi}$ and $\boldsymbol{\Phi}$ are the left and right singular vectors respectively, and $\boldsymbol{\Sigma}$ is the diagonal matrix of singular values. After that, the $n_\text{sensor}$ most dominant left singular vectors $\boldsymbol{\Psi}_{n_\text{sensor}}$ are extracted from the singular value decomposition. The final step involves computing the QR factorization with column pivoting of these vectors and selecting the top pivot locations as sensors.

\subsection*{Bayesian experimental design}
The Bayesian experimental design method \cite{huan2024optimal, verma2020} uses Bayesian inference to sequentially identify the design parameter $\boldsymbol{s}$ that provides the maximum information gain from prior to posterior, or equivalently the mutual information between the quantity of interest $\boldsymbol{q}$ for inference and the observable data $\boldsymbol{d}$. First, a uniform prior probability distribution $Pr(\boldsymbol{q})$ is constructed. According to Bayes' rule, the posterior probability distribution $Pr(\boldsymbol{q} \mid \boldsymbol{d},\boldsymbol{s})$ is proportional to the product of the prior distribution $Pr(\boldsymbol{q})$ and the likelihood $Pr(\boldsymbol{d} \mid \boldsymbol{q},\boldsymbol{s})$. The information gain is measured by a utility function $\mathcal{U}(\boldsymbol{s})$, defined based on the concept of relative entropy, also known as the Kullback–Leibler (KL) divergence, given by 
\begin{gather}
    \begin{aligned}
    \mathcal{U}(\boldsymbol{s}):=\mathbb{E}_{\boldsymbol{d} \mid \boldsymbol{s}}[KL(\boldsymbol{s},\boldsymbol{d})] & =\int_{\mathcal{H}} KL(\boldsymbol{s}, \boldsymbol{d}) Pr(\boldsymbol{d} \mid \boldsymbol{s}) \, \mathrm{d} \boldsymbol{d} \\
    & =\int_{\mathcal{H}} \int_{\mathcal{R}} Pr(\boldsymbol{q} \mid \boldsymbol{d}, \boldsymbol{s}) \ln \frac{Pr(\boldsymbol{q} \mid \boldsymbol{d}, \boldsymbol{s})}{Pr(\boldsymbol{q})} Pr(\boldsymbol{d} \mid \boldsymbol{s}) \, \mathrm{d} \boldsymbol{q} \, \mathrm{d} \boldsymbol{d}~.
    \end{aligned}
\end{gather}

We evaluate the utility function with the target quantity $\boldsymbol{q} = [q^{(1)}, \cdots, q^{(N_t)}]^T$ and the corresponding observable data for a candidate configuration as $\boldsymbol{d}$ (Eq.~\ref{equ:data_matrix}), where the specific configuration is considered to be the design parameter $\boldsymbol{s}$. To construct the likelihood model $Pr(\boldsymbol{d} \mid \boldsymbol{q})$, the continuous target quantity $\boldsymbol{q}$ is categorized into discrete bins based on its magnitude. Within each bin, the observable data is assumed to follow a Gaussian distribution, $Pr(\boldsymbol{d}_i \mid \boldsymbol{q}) \sim \mathcal{N}(\boldsymbol{\mu}_{i}, \boldsymbol{\Sigma}_{i})$, with the mean vectors and covariance matrices computed directly by fitting the available simulated data.

To mitigate the signal correlation issue, similar to the challenge previously addressed by \cite{attia2022optimal} in the context of measurement noise, we introduce a baseline level of independent, uncorrelated noise to the diagonal of the empirical covariance matrices: $\tilde{\boldsymbol{\Sigma}}_i = \boldsymbol{\Sigma}_i + \sigma_{noise}^2 \boldsymbol{I}$. It informs the method that adjacent sensors share highly correlated signals that can be easily dominated by their independent noise. The variance $\sigma_{noise}^2$ is tuned to ensure the algorithm spatially distributes the sensors and to maximize predictive performance. 

Since the $Pr(\boldsymbol{d} \mid \boldsymbol{s})$ cannot be evaluated analytically, evaluating the double integral requires a Nested Monte Carlo (NMC) estimator \cite{ryan2003estimating}. To accelerate computation, we employ Importance Sampling \cite{feng2019layered} for the inner evidence estimation. The utility for a sensor location $i$ is approximated as:
\begin{gather}
    \hat{\mathcal{U}}_i \approx \frac{1}{N_{MC, out}} \sum_{j=1}^{N_{MC, out}}\left\{\ln Pr\left(\boldsymbol{d}^{(j)}_i \mid q^{(j)}\right)-\ln \left[ \frac{1}{N_{MC, in}} \sum_{k=1}^{N_{MC, in}} \frac{Pr\left(\boldsymbol{d}^{(j)}_i \mid \tilde{q}^{(j,k)}\right) Pr(\tilde{q}^{(j,k)})}{Pr_{\text{prop}}(\tilde{q}^{(j,k)})} \right]\right\},
\end{gather}
where $N_{MC, out}$ and $N_{MC, in}$ are the number of outer and inner Monte Carlo samples, respectively. With the Importance Sampling technique, the proposal distribution $Pr_{\text{prop}}(\tilde{q})$ concentrates the inner samples around the corresponding outer parameter $q^{(j)}$ rather than sampling broadly from the uniform prior. 

To overcome the combinatorial complexity of sensor placement, we utilize a stochastic greedy algorithm \cite{mirzasoleiman2015lazier}. By evaluating the marginal utility, $i^*=\text{argmax} _{i \in \mathcal{V}_{sub}} \hat{\mathcal{U}}_i$, over only a randomized subsample $\mathcal{V}_{sub}$ of remaining candidates at each step, we reduce computational costs while maintaining a theoretical approximation guarantee. The sequential greedy algorithm allows any desired number of sensors to be extracted after a single execution.


\subsection*{Correlation-aware clustering: image clustering}
The images used in this section are sourced from the ImageNet dataset \cite{deng2009imagenet} and are downsampled to half of their original resolution for the naive FA demonstration and to one quarter for the clustering demonstration.

Clustering is performed using the AP algorithm, with parameters set to a damping factor of 0.9, a maximum of 10,000 iterations, 5 convergence iterations, and a preference of -40. The affinity metric for the clustering algorithm is computed as the Euclidean distance between each pixel, incorporating both the RGB color channels $R$, $G$, $B$, and the $x, y$ spatial coordinates, all scaled to the range [0, 1], given by
\begin{gather}
    \text{Affinity}={({\Delta R}^2+{\Delta G}^2+{\Delta B}^2+{\Delta x}^2+{\Delta y}^2)}^{\frac{1}{2}},
\end{gather}
where $\Delta$ denotes the difference between two pixels.

The naive FA is conducted by applying the IG algorithm to a pre-trained EfficientNet model \cite{tan2021efficientnetv2} for image classification.

\subsection*{Attribution based on correlation: correlated synthetic data}
The dataset generation follows a structured procedure to create variables with the desired correlations. First, inputs for Sensor 1 are randomly generated as an independent reference dataset. Next, inputs for Sensor 2 and Sensor 3 are generated by combining the Sensor 1 inputs with a weighted random component to achieve specified correlation coefficients with the Sensor 1 data. The input datasets are standardized to have a zero mean and unit variance. A dataset for the target variable $\boldsymbol{q}$ is then generated while ensuring specified correlations with the input datasets. The target dataset is constructed as a weighted linear combination of the input datasets, with weights determined by the desired correlation values. If necessary, an additional random component, orthogonal to all input datasets, is introduced to preserve unit variance while ensuring the specified correlations hold.

Clusters are produced using the AP algorithm with the affinity metric set to Pearson's correlation coefficient. The algorithm parameters are configured with a damping factor of 0.5, a maximum of 10,000 iterations, 10 convergence iterations, and a preference value of 0.7.

The IG algorithm for FA is performed on the MLP model trained with the datasets for the sensors and the target variable. The MLP architecture consists of a fully connected feedforward network with an input size equal to the number of sensors. It includes three hidden layers, each with 8 neurons, followed by batch normalization and LeakyReLU activation. The model was trained using the Adam optimizer with adaptive learning rate and $L_2$ regularization of 0.001 to prevent overfitting. The mean squared error loss function was used to measure the model’s performance.


\subsection*{Structural health monitoring}
\label{sec:SHM}

Inspired by Friswell and Castro-Triguero \cite{friswell2015} and Liu et al. \cite{liu2018optimal}, the beam consists of 30 candidate nodes with only a translational degree of freedom, meaning the sensor inputs are the deflections at each node. It has a length $L$ of 0.45 m, a width of 20 mm, and a thickness of 2 mm. The beam's elastic modulus is 32 GPa and density is 5219 kg/m$^3$. The first three vibration modes are considered. The mode shapes of the cantilever beam form a matrix $\Phi \in \mathbb{R}^{M \times N}$, where $M$ is the number of candidate sensor nodes, $N$ is the number of vibration modes. The mode shapes for a continuous cantilever beam are given by
\begin{equation}
    X_N(x) = \left[ \cosh(\beta_N x) - \cos(\beta_N x) \right] 
    - \sigma_N \left[ \sinh(\beta_N x) - \sin(\beta_N x) \right],
\end{equation}
where
\begin{equation}
    \sigma_N = \frac{\sinh(\beta_N L) - \sin(\beta_N L)}{\cosh(\beta_N L) + \cos(\beta_N L)},
\end{equation}
with $N=1,2,3$. $\beta_N L$ is obtained from the cantilever characteristic equation
\begin{equation}
    \cosh(\beta_N L)\cos(\beta_N L)+1=0.
\end{equation}
The mode shapes are normalized by their 2-norm and mass. For every beam profile, the deflections at the nodes are a linear superposition of the three mode shapes, as described by $u_s = \boldsymbol{\Phi}_s q$, where $u_s$ is a vector containing the beam deflections at each sensor node, $\boldsymbol{\Phi}_s\in \mathbb{R}^{n_\text{sensor} \times N}$ is the matrix containing the mode shapes reduced to the sensor nodes with $n_\text{sensor}$ denoting the desired number of optimal sensors, and $q$ refers to the modal coefficient vector. A dataset of 50,000 beam profiles is generated by uniformly sampling each modal coefficient from the interval $[-1,1]$.

The clustering implementation uses the same algorithm and parameters specified in Section \textit{Attribution based on correlation: correlated synthetic data}, with the preference parameter adjusted to 0.991 to yield 19 clusters. To implement CAAF, we construct a data-driven model consisting of an MLP that takes deflection measurements at cluster center nodes as inputs and predicts modal coefficients as outputs. While maintaining the basic architecture described in Section \textit{Attribution based on correlation: correlated synthetic data}, we modify the network by using ReLU activation functions in place of LeakyReLU and configuring each hidden layer with 12 neurons.

The main objective for finding OSP on a structure is to ensure that the mode shapes sampled at the sensor nodes are linearly independent and provide low-variance estimation of the structural response. Success in achieving these goals is commonly measured using the metrics RMS of the off-diagonal entries of the mass-weighted modal assurance criterion (MMAC) matrix, CN of the mode shape matrix, and DET of the reduced Fisher information matrix. RMS of the MMAC measures the level of orthogonality between the reduced mode shapes, where
\begin{equation}
    MMAC_{ij} = \frac{\left( \Phi_s^{(i)T} \mathbf{M}_s \Phi_s^{(j)} \right)^2}
{\left( \Phi_s^{(i)T} \mathbf{M}_s \Phi_s^{(i)} \right) 
\left( \Phi_s^{(j)T} \mathbf{M}_s \Phi_s^{(j)} \right)}
\end{equation}
for $i,j=1,2,\dots,N$, the superscript $^T$ denotes transpose, and $\mathbf{M}_s \in \mathbb{R}^{n_\text{sensor} \times n_\text{sensor}}$ is the mass matrix corresponding to the final sensor set; a lower RMS indicates a better modal orthogonality. CN of the mode shape matrix $\boldsymbol{\Phi}_s$ quantifies the linear independence between the mode shapes; a value closer to unity indicates better linear independence. DET of the reduced Fisher information matrix $ \boldsymbol{A}_s$, defined as $\boldsymbol{A}_s=\boldsymbol{\Phi}_s^T \boldsymbol{\Phi}_s$, reveals the level of information contained in the reduced mode shapes and variance of the estimated response; an optimal sensor configuration should maximize the DET.

We compare CAAF to two existing analytical methods for OSP: EI and KE. The EI method \cite{kammer1991sensor} optimizes sensor placement by maximizing the linear independence of target mode shapes. Given a mode shape matrix $\boldsymbol{\Phi} \in \mathbb{R}^{M \times N}$, the method evaluates the contribution of each sensor location through the Fisher information matrix $\boldsymbol{A} = \boldsymbol{\Phi}^T \boldsymbol{\Phi}$. The EI vector $\boldsymbol{E}_f \in \mathbb{R}^M$ is computed as the diagonal of $\boldsymbol{\Phi} \boldsymbol{A}^{-1} \boldsymbol{\Phi}^T$, where each entry quantifies a sensor's ability to preserve mode shape independence. The sensor corresponding to the lowest $\boldsymbol{E}_f$ entry is iteratively eliminated until the desired number remains, ensuring minimal information loss. Our EI sensors match those from Friswell and Castro-Triguero \cite{friswell2015}, which validates our implementation. The KE method \cite{kammer1991sensor} identifies OSP by maximizing the captured vibrational energy of target modes. For a structure with a complete mass matrix $\mathbf{M} \in \mathbb{R}^{M \times M}$, the energy metric $\boldsymbol{KE}$ is computed as
\begin{equation}
    \boldsymbol{KE} = \boldsymbol{\Phi} \otimes \mathbf{M}\boldsymbol{\Phi},
\end{equation}
 where $\otimes$ represents element-wise multiplication. Sensor locations are ranked by their row-summed modal energies, selecting positions with the highest dynamic activity.

 We also assessed the CAAF sensors against three brute-force benchmarks: Min RMS, Min CN, and Combined Score. These optimal sensors are identified by minimizing the RMS, CN, and the custom combined score defined in Equation \ref{combined_score}, respectively, via a brute-force search through all possible unordered combinations. To prevent one metric from dominating the selection process, the combined score acts as a dimensionless cost function that normalizes each metric against a predefined reference parameter. Specifically, the $RMS_{\text{ref}}$ is set to 0.01, the $CN_{\text{range}}$ is set to 0.5, and the $DET_{\text{max}}$ is defined as 3,700,000.

\subsection*{Airfoil lift prediction}
\label{airfoil_method}
The flow simulation uses a finite-volume, unstructured-mesh LES solver \cite{you2008discrete}. The spatially-filtered incompressible Navier-Stokes equations are solved with second-order accuracy employing cell-based, low-dissipative, and energy-conservative spatial discretization and a fully-implicit, fractional-step time-advancement method with the Crank–Nicolson scheme. The Poisson equation for pressure is solved using the algebraic multigrid method \cite{Ruge1987AMG}. The subgrid-scale stress is modeled using the dynamic Smagorinsky model \cite{germano1991dynamic,lilly1992proposed}.

\begin{figure}[t]
\centering
\includegraphics[width=0.7\textwidth]{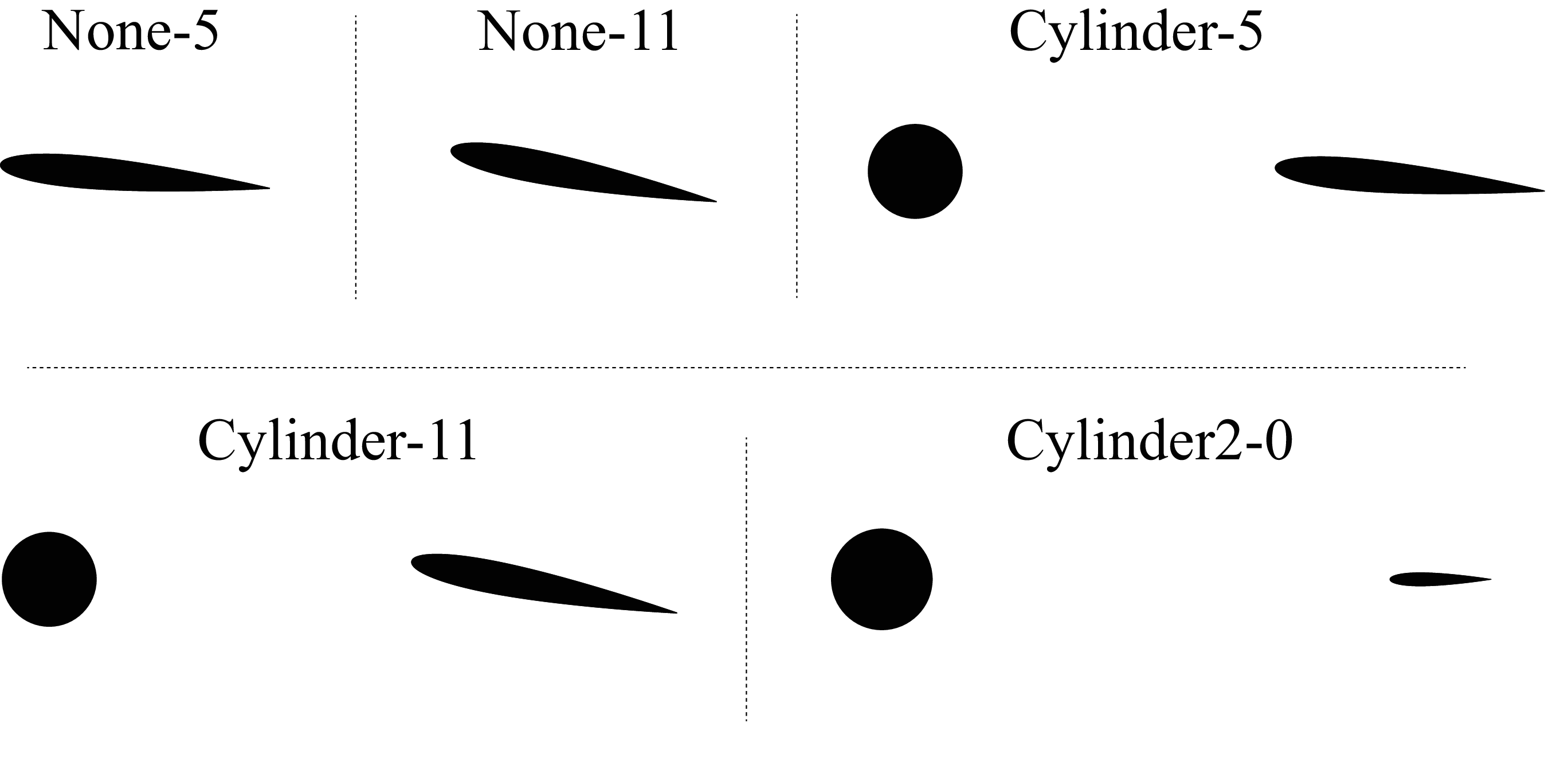}
\caption{\textbf{Schematics of simulated airfoil configurations.} 
Schematics of the geometries listed in Table~\ref{tab:cases}, featuring an airfoil at either 5 or 11 angle of attack. 
Configurations include those with gusty inflow generated by cylinders of two different sizes (``Cylinder-5'', ``Cylinder-11'', and ``Cylinder2-0'') and baseline cases without cylinders (``None-5'' and ``None-11''). 
The schematic for ``Cylinder2-0'' is scaled by a factor of 3/8 relative to the others.}
\label{fig:cases}
\end{figure}

Schematics of the simulation setup are shown in Fig.~\ref{fig:cases}. A NACA 0012 airfoil with chord length $C$ is placed horizontally at the origin of the coordinate system. Gusty inflows are generated by placing a cylinder directly upstream of the airfoil. Such geometry has been extensively studied and characterized both experimentally and numerically by others \cite{lefebvre2018,zhang2022,jiang2015}. Boundary conditions for both the airfoil surface and cylinder are defined as solid no-slip walls. The simulation domain is periodic in the spanwise direction with the domain size $L_z = 0.1C$. Uniform inflow of velocity $U_\infty$ is set to come in through the left and bottom surfaces of the domain at $Re_c = 10^4$ with an angle (equal to the AoA of the airfoil) relative to the $x$-axis. Additionally, convective outflow boundary conditions are set for both the top and right surfaces of the domain. The cylinder size, $D$, and its distance from the airfoil, $L_{c,a}$, vary across different cases. 

\begin{table}[h]
    \centering
    \caption{\textbf{Parameters of the simulated cases.}}
    \label{tab:cases}
    \begin{tabular}{lcccc}
    \toprule
        Case & Disturbance & AoA & $D/C$ & $L_{c,a}/D$\\
    \midrule
        None-5 & None & 5$^{\circ}$ & N/A & N/A\\
        None-11 & None & 11$^{\circ}$ & N/A & N/A\\
        Cylinder-5 & Cylinder wake & 5$^{\circ}$ & 0.35 & 3.8\\
        Cylinder-11 & Cylinder wake & 11$^{\circ}$ & 0.35 & 3.8\\
        Cylinder2-0 & Cylinder wake & 0$^{\circ}$ & 1.0 & 5.0\\
    \bottomrule
    \end{tabular}
    \caption*{{\footnotesize Type of upstream disturbance, the angle of attack (AoA) of the airfoil, the cylinder diameter relative to the chord ($D/C$), and the distance between the cylinder center and the airfoil leading edge ($L_{c,a}/D$) are listed for each case.}}
\end{table}
A total of five different cases are simulated, named ``None-5'', ``None-11'', ``Cylinder-5'', ``Cylinder-11'', and ``Cylinder2-0'', respectively. Schematics of the simulated geometry for these cases are displayed in Fig.~\ref{fig:cases}. 
Table~\ref{tab:cases} summarizes the parameters associated with the aforementioned cases.

The computational meshes used in the simulations consist of approximately 20 million cells. The airfoil is enclosed by a C-type mesh, which features approximately 400 cells along the airfoil's circumference. Approximately 160 cells are used to discretize the surface of the cylinder in cases where it is present. The boundary layer on the airfoil is captured by a mesh with $\Delta y_1/C \approx 0.0015$, where $\Delta y_1$ is the height of the first wall-normal cell. The spanwise direction is uniformly discretized into 128 cells. To verify the LES results, we compare the results of pressure coefficient $C_p$ and skin-friction coefficient $C_f$, defined as $C_p = P/\left(0.5 \rho U_\infty^2\right)$, and $C_f = \tau_w/ \left(0.5 \rho U_\infty^2\right)$
respectively, where $\rho$ is the fluid density and $P$ is the gauge pressure, around the airfoil from the ``None-5'' case with those generated from DNS by Gao et al. \cite{gao2019} at the same $Re_c$. The present LES results agree well with the reference DNS results. Additionally, a grid convergence test performed by repeating the ``None-11'' case with a finer mesh of 40 million cells produces less than 3\% difference in the time-averaged lift coefficients, suggesting that the current grid resolution is sufficiently high.

The lift force, $F_L$, is obtained by numerically integrating the surface pressure and shear stress components normal to the freestream. The lift coefficient is then calculated using the lift force by
\begin{gather}
    C_L = \frac{F_L}{\frac{1}{2} \rho U_\infty^2 L_zC}.
\end{gather} 
The pressure is scaled by $P/(\rho U_\infty^2)$. For additional details on the computational setup and validation results, readers are referred to our previous work \cite{leung2024}.

With data for the pressure distribution and the lift coefficients generated from the simulations, optimal sensor configurations are identified using CAAF and other data-driven methods: naive FA, Pivoted-QR, and Bayesian experimental design. Baseline configurations given by uniformly distributed sensors are also tested. The simulation data serve as inputs to the data-driven methods for identifying optimal sensor locations. While Pivoted-QR requires only pressure distribution data as input, Bayesian experimental design and FA methods require both pressure and lift data.

Neglecting the spanwise direction, the airfoil surface is spatially divided into a total of $M=376$ candidate pressure sensor locations following the computational grid, symmetrically distributed about the chord, covering both the suction and pressure sides. The pressure measurements are averaged over the spanwise direction. The five different cases mentioned above are grouped into training and testing sets. The training set comprises ``None-5'', ``None-11'', and ``Cylinder-5'' cases, whereas the testing set includes unseen data from all five cases. Only the data from the training cases are used in the data-driven methods to determine the optimal sensor locations with $N_t > 30,000$. The implementation steps for CAAF, Pivoted-QR, Bayesian experimental design, and uniform distribution methods are described as follows.

The clustering step in the CAAF is implemented by the AP algorithm with Pearson's correlation coefficient between the pressure inputs at different candidate sensor locations, $\boldsymbol{P}_i$, as the affinity metric. The damping and preference parameters are set to 0.5 and the median of the input similarities, respectively. The lift predictions are made by training MLP models with the instantaneous pressure measurements from the identified sensors as input and the lift coefficients at the same time step as output. The MLP model employs an architecture similar to that outlined in Section \textit{Attribution based on correlation: correlated synthetic data}, with the addition of an extra hidden layer. The pressure and lift coefficient data are interpolated to a uniform time grid with an interval of $0.005 C/U_\infty$. The training dataset is normalized and shuffled into batches of 64 data points. Normalization of the dataset is accomplished by scaling both the input pressure and output lift coefficients, \emph{i.e.} $C_{L,scaled}=(C_L-\langle C_L\rangle)/(\max(C_L)-\min(C_L))$. The averaged lift coefficients, $\langle C_L\rangle$, used for both normalization and prediction error calculations, are computed across the temporal snapshots. Normalization is performed separately for the data of each case. The averaged normalized prediction error, $\left\langle\varepsilon^*\right\rangle$, shown in Fig~\ref{fig:error_sensors}b, is calculated by normalizing the $L_2$ error of each method against the CAAF sensor baseline at each specific sensor count, making the $\left\langle\varepsilon^*\right\rangle$ identically equal to 1 for the CAAF sensors across all cases. This normalization procedure ensures a fair comparison across different sensor counts by standardizing the performance regardless of the number of sensors utilized.

To identify the Pivoted-QR sensors, the time series pressure data from the simulation are arranged into a matrix according to Eq.~\ref{equ:data_matrix}. The sensor selection then follows the algorithm described in the \textit{Pivoted-QR} section.

For the Bayesian experimental design method, the lift coefficients $\boldsymbol{C_L} = [C_L^{(1)}, \dots, C_L^{(N_t)}]^T$ represent the target quantity $\boldsymbol{q}$ while the pressure $\boldsymbol{P}$ serves as the observable data $\boldsymbol{d}$. We select 18 uniformly spaced bins across the $C_L$ range with over 1,500 samples per bin. The added sensor noise variance $\sigma_{noise}^2$ is 0.001. Sample sizes for the NMC are tuned for convergence at $N_{MC, out} = 30,000$ and $N_{MC, in} = 5,000$.


Finally, the baseline sensor configurations are obtained by uniformly distributing the sensors on the surface of the airfoil, resulting in an equally spaced sensor array. Moreover, to ensure the information from both the leading and trailing edges is captured by the baseline configurations, two sensors are fixed at the two edges before evenly placing the rest across the airfoil surface.


\subsection*{Turbulent wall-normal velocity estimation}
In order to obtain high-fidelity data for wall-normal velocity and wall pressure, we perform a DNS of an incompressible turbulent channel flow at \(Re_{\tau} \approx 186\). The simulations are performed by discretizing the incompressible Navier–Stokes equations with a staggered, second-order-accurate, central finite-difference method in space \cite{orlandi2000}, and an explicit third-order-accurate Runge–Kutta method for time advancement \cite{wray1990}. The system of equations is solved via an operator splitting approach \cite{chorin1968}. Periodic boundary conditions are imposed in the streamwise and spanwise directions, and the no-slip condition is applied at the walls. The code has been validated in previous studies of turbulent channel flows \cite{bae2018, bae2019}. The streamwise, wall-normal and spanwise domain sizes are \(L_{x,c}^+ \approx 5,300\), \(L_{y,c}^+ \approx 372\) and \(L_{z,c}^+ \approx 2,000\), respectively. The grid spacings in the streamwise and spanwise directions are uniform, with \(\Delta x^+, \Delta z^+\approx 5.2\); non-uniform meshes are used in the wall-normal direction, with the grid stretched towards the wall according to a hyperbolic tangent distribution with \(\min(\Delta y^+) \approx 0.18\) and \(\max(\Delta y^+) \approx 7.5\). The dataset comprises $N_t = 900$ temporal snapshots with $\Delta t^+ = 1.06$, each divided into 57 sensing sections, yielding over 50,000 data points scaled to the range [-1,1]. The sensing plane contains $M=361$ candidate sensor locations. The testing dataset consists of all snapshots from one of the sensing sections.

We applied AP clustering to the candidate sensor locations based on wall-pressure correlations, with damping and preference parameters set to 0.5 and 0.9, respectively. All other hyperparameters maintain the same configuration detailed in Section \textit{Attribution based on correlation: correlated synthetic data}.
We then train a TCN model to predict $v_{10}$ at the center of the sensing plane using $p_w$ measurements from the 58 cluster centers identified by AP. The TCN is well-suited for short-term temporal prediction using spatially distributed measurements, making it an ideal choice for the present study. The TCN architecture consists of two stacked 1D convolutional layers (kernel size 3, zero-padding) with ReLU activations to extract local temporal dependencies. The resulting multi-channel feature maps are then flattened and mapped to a continuous scalar prediction via a final fully connected layer. It utilizes an input sequence length of 10 and a hidden dimension of 64. The model was trained for 150 epochs by minimizing the mean squared error loss. Optimization was performed using the Adam algorithm with an adaptive learning rate initialized to 0.0001 and a weight decay of 1e-5. The model achieves a correlation of 0.953 between prediction and reference with 58 cluster centers as inputs. The correlation is computed using the Pearson correlation coefficient. The averaged wall-normal velocity, $\langle v_{10} \rangle$, used in the prediction error calculation is computed across the temporal and homogeneous directions. This performance is comparable to the results reported by Park and Choi \cite{park2020machine}, who achieved a maximum prediction correlation of 0.96 using the full-state wall pressure as input.

To identify the Pivoted-QR sensors, the time series wall pressure data from the simulation are arranged into a matrix according to Eq.~\ref{equ:data_matrix}. The sensor selection again follows the algorithm described in the \textit{Pivoted-QR} section.

For the Bayesian experimental design method, the wall-normal velocity $\boldsymbol{v}_{10} = [v_{10}^{(1)}, \dots, v_{10}^{(N_t)}]^T$ represents the target quantity $\boldsymbol{q}$ while the wall pressure $\boldsymbol{P_w}$ serves as the observable data $\boldsymbol{d}$. We select 25 uniformly spaced bins across the $v_{10}$ range with over 2,000 samples per bin. The added sensor noise variance $\sigma_{noise}^2$ is 0.005. Sample sizes for the NMC are tuned for convergence at $N_{MC, out} = 20,000$ and $N_{MC, in} = 4,000$.

\section*{Data Availability}

 The datasets used in this study are publicly available in the GitHub repository \cite{leung2026github}. The channel flow velocity and wall pressure data for the wall-normal velocity estimation are accessible via Figshare \cite{Leung2026_figshare}.

\section*{Code Availability}
The source code used to conduct the experiments and analyze the data is publicly available in the GitHub repository \cite{leung2026github}.

\section*{Acknowledgments}
Not applicable

\bibliography{Ref}

\section*{Funding Statements}

We gratefully acknowledge the funding provided for this research by the Keck Scholar-Fellow Bridge Initiative from the W. M. Keck Foundation, the SciAI Center, supported by the Office of Naval Research (ONR) under grant number N00014-23-1-2729, the Carver Mead New Adventures Fund, and the Center for Autonomous Systems and Technologies at the California Institute of Technology. Computational time was provided by the Discover project at Pittsburgh Supercomputing Center through allocation PHY240020 from the Advanced Cyberinfrastructure Coordination Ecosystem: Services \& Support (ACCESS) program, which is supported by NSF grants No.~2138259, No.~2138286, No.~2138307, No.~2137603, and No.~2138296.

\section*{Author Contributions Statement}

H.J.B. conceived the experiments and provided guidance;  S.C.L. developed the framework, conducted the experiments, analyzed the results, and wrote the manuscript; D.Z. assisted in conducting the experiments and the analysis. All authors reviewed the manuscript. 

\section*{Competing Interests}
The authors declare no competing interests.



\end{document}